\documentclass{elsart1p}

\providecommand{\openone}{\leavevmode\hbox{\large1\kern-7.3pt\normalsize1}}
\newcommand{\be}{\begin{equation}}
\newcommand{\ee}{\end{equation}}
\newcommand{\ba}{\begin{eqnarray}}
\newcommand{\ea}{\end{eqnarray}}

\newcommand{\fr}[2]{{\frac{#1}{#2}\,}}

\renewcommand{\ln}{{\rm ln}}

\newcommand{\bi}{\begin{itemize}}
\newcommand{\ei}{\end{itemize}}

 \usepackage{graphicx}

\usepackage{amssymb}

\begin{document}

\begin{frontmatter}



\title{Equation of state of zero-temperature quark matter with finite quark masses}


\author{Aleksi Vuorinen}

\address{CERN, Physics Department, TH Unit, CH-1211 Geneva 23, Switzerland\\
   Institut f\"ur Theoretische Physik, TU Wien, Wiedner Hauptstr. 8-10, A-1040 Vienna, Austria}

\begin{abstract}
We outline the key elements of a recent calculation aimed at determining the equation of state of deconfined (but unpaired) quark matter at zero temperature and high density, using finite quark masses. The computation is performed in perturbation theory up to three loops, and necessitates the development and application of some novel computational tools. In this talk, we introduce the basic features of these new techniques and review the main sources of motivation for considering finite quark mass effects in perturbation theory.
\end{abstract}

\end{frontmatter}

\section{Introduction}

The precise determination of the equation of state of deconfined quark matter is one of the fundamental challenges in finite-temperature QCD, and a considerable amount of effort has been invested in it. An approach complementary to lattice QCD is to apply perturbation theory, where one starts from the regime of asymptotically high temperatures and/or densities, in which the coupling is known to be vanishingly small, and attempts to extrapolate the results to intermediate couplings. At high temperatures $T$ and quark chemical potentials $\mu$ not exceeding $T/g$, the weak coupling expansion of the pressure has been extended to four-loop order, or up to and including the ${\mathcal O}(g^6\ln\,g)$ term, and has been demonstrated to provide reasonable agreement with lattice results down to temperatures of the order of 5$T_c$ \cite{klrs,av}. At zero temperature, as well as for $0<T\lesssim g\mu$, the expansion is however only known to order $g^4$ \cite{fmcl,ikrv}, and has furthermore only limited physical relevance due to the pairing instability leading to color superconductivity \cite{arw1}, which is a fundamentally non-perturbative effect.

One aspect of the perturbative calculations that until recently has received much less attention is the role of finite quark masses, which in principle might play an important role at high, but not asymptotic energy densities. The only studies incorporating finite quark mass effects are at present at two-loop order \cite{ls,fr}, and while at high temperatures it appears that to a very good approximation one can obtain the quark mass dependence of the pressure from the free theory result \cite{ls}, things are drastically different close to zero temperature \cite{fr}. In this regime, the physical systems of interest are quark stars, the mass-radius relationship of which is strongly dependent on the equation of state of cold and dense quark matter. In addition, the quark mass dependence of the equation of state of zero-temperature quark matter is an essential ingredient in determining the mismatch of the Fermi surfaces of the up, down and strange quarks as a function of the quark chemical potential, which can be used to estimate the critical chemical potential for the breakdown of the color-flavor-locked (CFL) color superconducting phase \cite{arw2}.

In this talk, we wish to outline a new perturbative computation of the equation of state of zero-temperature quark matter to three loops, in which we assume to be dealing with two massless (up and down) and one massive (strange) quark flavor \cite{krsv}. This is the first example of a thermodynamical calculation incorporating finite masses at such a high perturbative order, and we are thus faced with the necessity to develop some novel computational machinery. We outline the general structure of the calculation in Section 2, and in addition introduce these new techniques. In Section 3, we then explain, how we intend to use the results of our (at the moment still unfinished) computation.

\section{Outline of the calculation}

As in any perturbative calculation, the evaluation of the three-loop pressure of zero-temperature quark matter involves three steps: First, identifying the graphs that contribute at the order one is investigating; second, reducing the necessary diagrams to a finite set of master integrals; and third, solving for these masters. In the case at hand, the first step has been solved long ago, and there in fact is no qualitative difference to the massless case considered already in Ref.~\cite{fmcl}: To obtain the pressure up to and including order $g^4$, one simply needs to compute all two-gluon-irreducible (2GI) graphs to three-loop order, and in addition perform the so-called plasmon sum of ring diagrams. As the latter part of the calculation turns out to be highly analogous to the massless case and presents few new computational challenges, in this presentation we focus on the rest, \textit{i.e.~}the two- and three-loop 2GI graphs displayed in Fig.~1.

\begin{figure}[t]
\begin{center}
\includegraphics*[width = 0.75\textwidth]{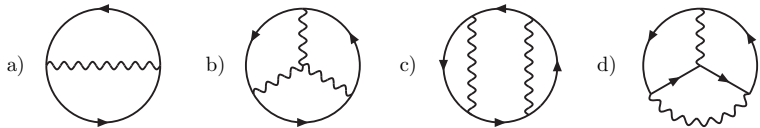}
\caption{The two- and three-loop 2GI diagrams contributing to the pressure.}
\end{center}
\end{figure}

%
%
%

For the 2GI diagrams, the procedure we follow consists of four parts: Performing the Lorentz contraction of indices to reduce the graphs to 'scalar' integrals, implementing so-called cutting rules on these scalars to reduce the four-dimensional integrals into three-dimensional ones, performing renormalization, and finally numerically evaluating the $3d$ integrals. Here, the cutting rules are simply a manifestation of the straightforwardly verifiable fact that a zero-temperature fermionic propagator with mass $m$ and chemical potential $\mu$ can in diagrammatic calculations be replaced by a 'shifted' propagator according to
\ba
\fr{1}{(p_0+i\mu)^2+\mbox{\boldmath$p$}^2+m^2} &\rightarrow & \fr{1}{p_0^2+\mbox{\boldmath$p$}^2+m^2}-i\delta(p_0^2+\mbox{\boldmath$p$}^2+m^2)\,\theta_{\mu,\, p_0},\\
\theta(\mu-{\rm Im}\, p_0)\,\theta({\rm Im}\, p_0)&\equiv& \theta_{\mu,\, p_0}.
\ea
Each scalar graph with $n$ loops can thus clearly be written as the sum of $n+1$ parts, which have been obtained by 'cutting' some number (0 to $n$) of the fermionic lines, corresponding to the number that the above delta function piece appears in it. In the cutting procedure, one replaces for the cut lines
\ba
p_0 &\rightarrow& i E({\mbox{\boldmath$p$}}), \label{onshell}\\
\int {\rm d}p_0&\rightarrow& -\fr{\theta(\mu-E({\mbox{\boldmath$p$}}))}{2E({\mbox{\boldmath$p$}})}, \label{p0rule}
\ea
while the uncut fermion lines no longer carry a chemical potential $\mu$ but naturally remain massive, and the bosonic lines stay fully intact. The possibly intricate issue related to squared propagators in this type of formalism (see \textit{e.g.~}the discussion in Sec.~3.7 of Ref.~\cite{kap}) is avoided by writing all squared propagators in terms of mass derivatives.

\begin{figure}[t]
\begin{center}
\includegraphics*[width = 0.75\textwidth]{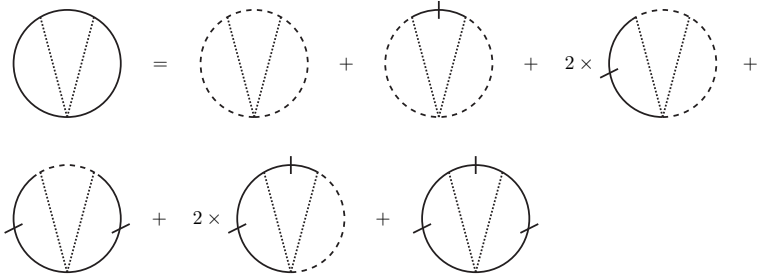}
\caption{A demonstration of the application of the diagrammatic cutting rules. The solid lines that have been marked with the short straight line have been cut, and the dashed and dotted lines stand for massive and massless scalar propagators ($\mu=0$ everywhere), respectively.}
\end{center}
\end{figure}
%

The application of the cutting rules is demonstrated for a generic three-loop scalar graph in Fig.~2. Most importantly, we note that on the right hand side of this equation, the only dependence on chemical potentials resides in the integration measure of the three-dimensional integrals corresponding to the cut lines. The integrands in these integrations are four-dimensional 0-, 2-, 4- and 6-point functions, where the external momenta have been put on shell according to the cutting rules laid out above and $\mu$ has been set to zero. This amounts to an enormous simplification in the calculations, as for zero chemical potential and temperature, one has the vast machinery of $T=0$ diagrammatic techniques, such as integration-by-parts identities, at hand. In practise, one may automatize the application of the cutting rules and IBP-identities to such an extent that the only manual labor necessary is in finding the results of the various master integrals from the literature. In this process, we have found the new Mathematica-code FIRE \cite{fire} immensely useful.

Finally, even performing renormalization is rendered quite straightforward in the approach described above. All UV singularities are present only in the $\mu$-independent integrands of the three-dimensional integrals, and can thus be removed before the (numerical) evaluation of the $3d$ integrals. These integrals are on the other hand at worst six-dimensional (three radial and three angular coordinates, corresponding to three $3d$ momenta and their relative angles), and can be handled straightforwardly with standard numerical tools.

\section{Applications}

Finally, let us outline the main applications of the (currently still ongoing) calculation described above. The motivation for the computation of the strange quark mass effects on the equation of state of cold quark matter is two-fold, on one hand originating from direct astrophysical applications, and on the other hand from an indirect use in the determination of the breakdown density of the color-flavor locked phase in color superconductivity.

The structure of quark stars is usually described through a numerical solution of the so-called Tolman-Oppenheimer-Volkov (TOV) equations, which require the equation of state of electrically neutral quark matter as a function of density as an ingredient. To first approximation, one may neglect the effects of color superconductivity in the bulk thermodynamic quantities, but as described in Ref.~\cite{fr}, the effects originating from the running of the strong coupling constant and the quark masses are at two-loop level still considerable, thus motivating the extension of the weak coupling expansion of the pressure to three loop order.

On the other hand, a \textit{necessary} (though not sufficient) criterion for the CFL phase to be realized in nature at a given density is that it is preferred with respect to ordinary, unpaired quark matter. Whether this is the case can be found by computing the splitting between the $u$, $d$ and $s$ quark Fermi momenta, and comparing this to the CFL gap. Again, previous studies have found the renormalization scale dependence of the two-loop perturbative result for the former to be too high for drawing any quantitative conclusions \cite{krishna}, so a three-loop evaluation of the quantity (necessitating the computation of the strange quark mass effects on the equation of state) is required.



\end{document}